\documentclass[preprint,aps,showpacs]{revtex4}
\usepackage{graphicx}

\begin{document}
\title{How ripples turn into dots: modeling ion-beam erosion under oblique incidence}
\author{Sebastian Vogel} \author{Stefan J. Linz}
\affiliation{Institut f\"ur Theoretische Physik, Universit\"at M\"unster, Wilhelm-Klemm-Str.9, 48149 M\"unster, Germany}
\date{\today}


\begin{abstract}
Pattern formation on semiconductor surfaces induced by low energetic ion-beam 
erosion under normal and oblique incidence is theoretically investigated 
using a continuum model in form of a stochastic, nonlocal, anisotropic 
Kuramoto-Sivashinsky equation. Depending on the size of the parameters 
this model exhibits hexagonally ordered dot, ripple, less regular 
and even rather smooth patterns. We investigate the transitional behavior  
between such states and suggest how transitions can be experimentally 
detected.  
\end{abstract}
\maketitle
Self-organized structure formation on the nanoscale induced by 
ion-beam erosion, i.e. the removal of target material by bombarding 
its surface with ionized particles, has recently turned into a highly 
active research area of surface science with deep connections to the
modern theory of non-equilibrium systems. Since the work of Navez et al. 
\cite{navez} in 1962 it is known that, under {\it oblique incidence} 
of the ions, often washboard-like {\it ripple} patterns  oriented normally 
to the beam direction  can be observed on the surface 
(for recent reviews cf. \cite{ibe-rev,makeev}). For {\it normal 
incidence} of the ions, however, just a sort of rough, unstructured surface 
evolution had been expected. So, it came as a surprise when Facsko et al. 
\cite{facsko1} observed quite regularly ordered, hexagonally 
arranged {\it dot} structures for semiconducting GaSb under 
{\it normal incidence} of low energetic Ar$^+$-ions.    
Parallel to that, Frost et al. \cite{frost1} found for rotated InP, InAs, 
InSb and GaSb targets under oblique ion incidence a variety of distinct 
patterns such as dot or square structures and even rather flat or smooth 
surfaces. Rather recently, experiments under {\it oblique incidence without 
target rotation} have revealed an even more puzzling picture.  
Using Si targets and ion energies $\leq 2000eV$, Ziberi et al. \cite{ziberi} experimentally 
detected transitions from ripples oriented {normally} to the beam direction to rather
smooth surfaces by increasing the incidence angle. Similar results have also been 
obtained by Ziberi et al. \cite{ziberi2}, for Ge and Si targets, where in the latter 
case almost perfectly straight ripples have been identified. 
Finally using GaSb targets and low ion fluxes, Allmers et al. \cite{allmers} identified 
a transition from hexagonal dot patterns to ripples oriented {tangentially} to 
the beam direction for small off-normal incidence.
\\
The theoretical understanding of the hexagonally ordered dot
structures under {\it normal incidence} has posed a particular challenge  
because  the standard continuum model for the evolution of the morphology
$H({\bf x},t)$ for ion-beam erosion, the anisotropic Kuramoto-Sivashinky 
equation (aKSE)\cite{bara,makeev,rost} does not seem to reproduce 
such patterns in its 
isotropic limit (iKSE), $\partial_t H = F_0+a_1\nabla^2 H + a_2\nabla^4 H + 
a_3( \nabla H)^2 + \eta$. Therefore, two distinct generalizations of
the iKSE have been put forward: Using different physical reasoning, 
Kim et al. \cite{kim} and Castro et al. \cite{castro} have suggested the 
inclusion of a term proportional to $\nabla^2(\nabla H)^2$. Although this 
extended KSE (already studied in the context of amorphous surface 
growth \cite{linz1}) might show patterns resembling hexagonally 
short-range ordered dots for some parameter
values \cite{castro}, it typically exhibits an irregular cellular 
pattern with remedies of a hexagonal arrangement \cite{linz1}. 
As an alternative to explain the hexagonally ordered dots, Facsko et al. 
\cite{facsko-t} suggested the inclusion of a damping term $bH$ with 
$b<0$ in the iKSE that is known to lead to hexagonal structures. This term,
physically interpreted as the effect of redeposition of sputtered 
particles, fails to fulfill the fundamental symmetry of translation 
invariance in erosion direction $H\rightarrow H+z$ with $z=$constant. 
To fix that, Facsko et al. suggested to replace $bH$ by $b(H-\overline{H})$
with $\overline{H}$ being the spatial average over some sample area.
Interpreting $\overline{H}$ as the spatial average over the whole sample, 
we showed in \cite{voli1} that such a non-local iKSE can be rigorously 
rewritten as a (local) damped iKSE (idKSE) by means of a temporally nonlocal
transformation and, therefore, put the applicability of the idKSE on 
solid theoretical grounds.\\ 
As a consequence of \cite{facsko-t, voli1}, several important questions arise: 
(i) How does a minimal model for ion-beam erosion under normal {\it and} 
oblique angle of incidence look like and what are its eminent emergent 
patterns? 
(ii) How does the transition from dots to ripples with varying angle 
of inclination happen? Is it a gradual change or a bifurcation?   
Does it occur at a zero or non-zero angle of incidence? 
(iii) How do experimentally observed smooth patterns 
with very low surface roughness fit in such a description? 
(iv) How do generic transitions between distinct patterns as 
function of the angle of incidence look like? 
In this letter, we theoretically investigate these questions on the 
basis of an anisotropically generalized version of a nonlocal KS 
equation and its damped counterpart.

{\it Model equation. -} Generalizing the recipe in \cite{voli1},  
the balance equation for the evolution of the height function $H(x,y,t)$, 
$\partial_t H={\nabla}\cdot{\bf J}_H + F +\eta$,  is determined by 
(i) relaxational currents caused by surface diffusion and assumed 
to be dominantly isotropic modeled by  ${\nabla}\cdot{\bf J}_H=a_2\nabla^4 H$ 
with $a_2<0$ and (ii) and detachment contributions $F=F[\partial_x H,\partial_y H,
H-\overline{H}]$ that depend on partial derivatives of $H$, $H-\overline{H}$ 
and all admissible combinations of them. Possible anisotropies stemming from ion-induced
surface diffusion are at this stage neglected. Orienting the coordinate system 
such that the tangential component of the incoming ion flux is parallel 
to the $x$-axis, any term depending on odd order derivatives with respect 
to $y$ is excluded due to the reflectional invariance along the $x$-axis. 
Expanding $F$ in its arguments and keeping solely lowest order terms yields
\begin{eqnarray}
\partial_t H&=&F_0 + b (H-\overline H)+a_0 \partial_x H 
               + (a_{1x} \partial^2_x + a_{1y} \partial^2_y) H +\nonumber \\
	    & &+ a_2 \nabla^4H + a_{3x}(\partial_x H)^2 + a_{3y}(\partial_y H)^2 + \eta.
\label{adksnl}
\end{eqnarray}
The physical significance of the terms on the rhs of (\ref{adksnl}) are constant erosion velocity $F_0$ if all other terms on the rhs
were zero, surface drift due to oblique incidence ($a_0$-term), 
redeposition ($b$-term), anisotropic surface roughening via 
Bradley-Harper mechanism \cite{bradley} 
($a_{1x},a_{1y}$ terms), isotropic thermal surface 
diffusion ($a_2$-term), anisotropic tilt-dependent sputter yield 
($a_{3x},a_{3y}$ terms) and isotropic stochasticity in 
the erosion process with Gaussian white noise $\eta(x,y,t)$ of  
covariance $2D$. 

Next we simplify (\ref{adksnl}).  
First, the drift term $a_0 \partial_x H$ in (\ref{adksnl}) is eliminated 
via the  transformation $H(x,y,t) \longrightarrow H(x-a_0 t, y, t)$ 
implying a motion of the resulting pattern with a constant speed $-a_0$ 
in the $x$ direction without any change of shape of the morphology of $H$. 
Second, (\ref{adksnl}) is recast to a local field
equation via the temporally nonlocal transformation \cite{voli1}
\begin{equation}
h =  H - \overline{H} + ({F_0}/{b})(1-e^{bt}) + e^{bt}\int_0^{t'}\partial_{t}\overline{H}e^{-bt'} dt'
\end{equation}
yielding
\begin{equation}
\partial_t h = b h+(a_{1x} \partial^2_x + a_{1y} \partial^2_y) h + a_2 \nabla^4h + a_{3x}(\partial_x h)^2 + a_{3y}(\partial_y h)^2 + \eta.
\label{adksnn}
\end{equation}
Third, rescaling the  
time by $-(a_{1x}^2/a_2)t\rightarrow t$,  
length scales by $\sqrt{a_{1x}/a_2}{\bf x}\rightarrow {\bf x}$,  
height by $-(a_{3x}/a_{1x})h\rightarrow h$ and noise by 
$(a_2a_{3x}/a_{1x}^3) \eta\rightarrow \eta$  
and introducing the three coefficients  
$\gamma=(a_2/a_{1x}^2)b>0$, 
$\alpha=a_{1y}/a_{1x}$, 
$\beta=a_{3y}/a_{3x}$, leads to 
\begin{equation}
\partial_t h= -(\gamma + \partial^2_x  + \alpha\, \partial^2_y  + \nabla^4) h  
	 + (\partial_x h)^2 + \beta(\partial_y h)^2 + 
	\eta. 
	\label{adks1} 
\end{equation}
The anisotropic, damped KS equation (adKSE) (\ref{adks1}) 
constitutes our minimal model for erosion under normal and oblique incidence. 
The parameters $\alpha$ and  $\beta$ measure the relative 
anisotropies of the surface roughening and the sputter yield whereas 
the third parameter $\gamma \geq 0$ the rescaled damping. The limit $\gamma=0$ 
is the aKSE \cite{makeev,bara,rost}, whereas the limit
$\alpha=1=\beta$ yields the idKSE \cite{facsko-t,voli1}.\\ 
Two properties of Eq.(\ref{adks1}) are of subsequent interest. 
(A) The parameter range $\alpha>1$ can be mapped to the range $0<\alpha< 1$ 
and vice versa by flipping the coordinate system, i.e. $x\rightarrow y$,
 $y\rightarrow x$ and  
rescaling time by $\alpha^2 t\rightarrow t$, 
length scales by $\sqrt{\alpha} {\bf x}\rightarrow {\bf x}$, 
height by $(\beta/\alpha)h\rightarrow h$ and the noise by 
$(\beta/\alpha^3) \eta\rightarrow \eta$ and transforming the coefficents, 
$\tilde{\gamma}=\gamma/\alpha^2$,
$\tilde{\alpha}=1/\alpha$ and 
$\tilde{\beta}=1/\beta$. This yields the functionally equivalent form  
$\partial_t h= -(\tilde{\gamma} +\partial^2_x  + \tilde{\alpha}\,\partial^2_y  + \nabla^4) h  
	 + (\partial_x h)^2 +\tilde{\beta}(\partial_y h)^2 + 
	\eta$. 
Consequently, the parameter sets $(\gamma,\alpha,\beta)$ with $\alpha<1$ and 
 $(\gamma/\alpha^2,1/\alpha,1/\beta)$ with $1/\alpha=\tilde{\alpha}\ge 1$ 
lead to the same dynamics.
(B) The unscaled version of  Eq.(\ref{adks1}), i.e. Eq.(\ref{adksnn}), 
possesses the invariance under the combined transformation 
$h\rightarrow -h, a_{3x}\rightarrow -a_{3x}, a_{3y}\rightarrow -a_{3y}$. This implies 
that the corresponding morphologies are reflected about $h=0$ if the signs of $a_{3x}$ 
and $a_{3x}$ are inverted, and, consequently, the role of mounds 
and valleys of the surface profile are exchanged. In Eq.(\ref{adks1}) this invariance 
is hidden in the scaling.\\
Since the coefficients in 
(\ref{adksnl}) depend on the angle of incidence and various other parameters 
such as ion flux and type, target material etc. and are not well known, we
subsequently study how the properties of (\ref{adks1}) parametrically depend on the
anisotropies and the damping without reference to a specific material and, by that, 
explore the pattern forming properties of Eq.(\ref{adks1}). 
\begin{figure}[t,b]
\centering
\includegraphics[width=.5\columnwidth]{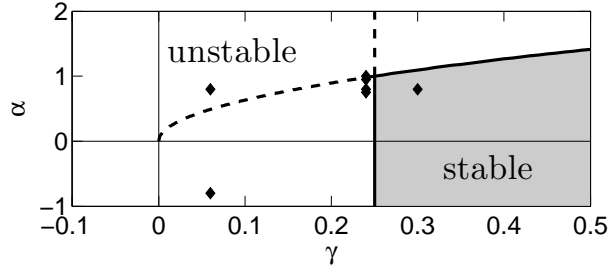}
\caption{Stability diagram of the flat front solution $h_{FF}=0$ for $\eta=0$. 
Markers indicate parameter values for subsequently used plots of the
morphology.}
\label{fig1}
\end{figure}

{\it Onset of pattern formation. -} The non-stochastic limit of (\ref{adks1}) 
obviously possesses a flat front solution $h_{FF}=h({\bf x},t)=0$ corresponding to 
$H_{FF}=F_0 t$ as primary pattern. Its linear stability, determined by the 
growth or decay of small superimposed perturbations $h({\bf x},t) \propto {\rm
exp}(i{\bf k}\cdot{\bf x}+\sigma t)$ in the linearized version of (\ref{adks1}) 
yields the growth rate $ \sigma({\bf k})=-\gamma+k_x^2+\alpha k_y^2-{\bf k}^4$ 
as function of the wave-number ${\bf k}=(k_x,k_y)$. The corresponding critical 
wave numbers are determined by (i) $k_x=0$ and $k_y=\sqrt{\alpha/2}$ or 
(ii) $k_x=\sqrt{1/2}$ 
and $k_y=0$. Consequently, $h({\bf x},t)=0$ is linearly unstable (i) with respect  
to plane wave perturbations in $y$-direction if  $\gamma<\alpha^2/4$ 
and $\alpha\geq1$ or (ii) with respect to plane wave perturbations in 
$x$-direction if $\gamma<1/4$ and $\alpha\leq 1$. In Fig. \ref{fig1} the 
results of the linear stability analysis of $h_{FF}=0$ are shown. 
Note the significant difference to the  aKSE 
corresponding to the vertical line at $\gamma=0$ where for $a_{1x}<0$ and/or
$a_{1y}<0$ no stable flat front solutions can be obtained. Similar arguments also 
apply to any nonlinear extension of the aKSE.

{\it Instability of perfect ripples. -}   
 Assuming $\eta=0$ and a perfect ripple state $h_n(x,y,t)=h_n(x,t)$ 
{\it normal} to the incidence direction ({\it n-ripples}) obeying       
$\partial_t h_n = - (\gamma+\partial^2_x+\partial^4_x) h_n 
+ (\partial_x h_n)^2$, the evolution of a slightly perturbed 
height function $h(x,y,t)=h_n (x,t)+ p(y,t)$ in $y$-direction is 
governed by  $\partial_t p = - (\gamma+\alpha\partial^2_y+\partial^4_y)p 
+\beta (\partial_y p)^2$. Linearizing and inserting periodic perturbations 
$p=\mathrm{e}^{i\tilde{k}y+\tilde{\sigma} t}$ yields 
the maximum growth rate $\tilde{\sigma}_{max}=-\gamma+\alpha^2/4$ 
implying that n-ripple patterns are unstable against $p$-perturbation for 
$\gamma<\alpha^2/4$ with a critical wave number 
$\tilde{k}=\sqrt{\alpha/2}$. 
Similarly, a perfect ripple state $h_t(x,y,t)=h_t(y,t)$ 
{\it tangential} to the incidence direction ({\it t-ripples}) 
is unstable with respect to a slight perturbation $p(x,t)$ in $x$-direction 
for $\gamma<1/4$ with a critical wave number 
$\tilde{k}=\sqrt{1/2}$. Consequently, such perfect t- or n-ripple
patterns are always unstable in the aKSE. In the adKSE, however, they are, at least 
for these specific types of perturbations, stable in some parameter ranges. 
Since Ziberi et al. \cite{ziberi2} were able to experimentally produce basically 
perfectly straight ripples, the necessity of the damping term in our 
model equation seems to stringent.
\begin{figure}[tb]
\centering
\includegraphics[width=.8\textwidth]{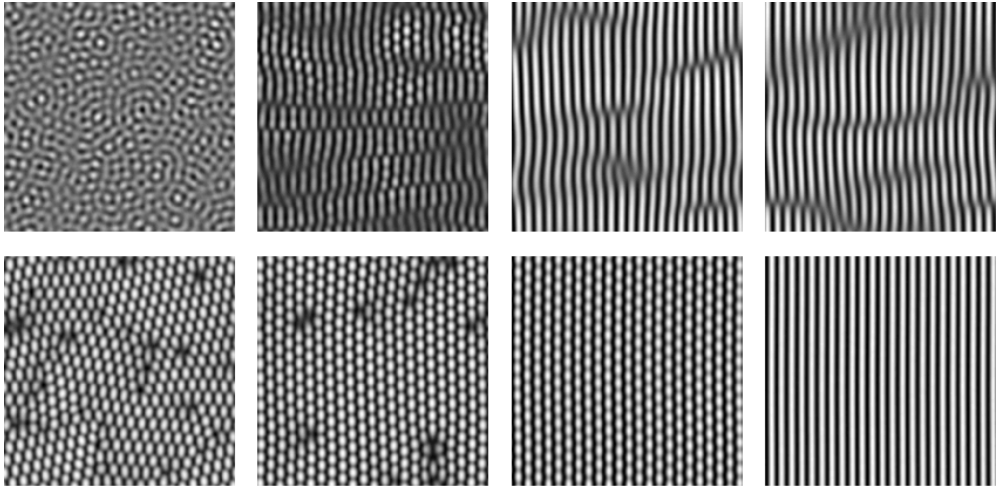}
\caption{Snapshot of the surface morphologies for $\gamma=0.24$, $\beta=2.0$ 
each and $\alpha=1.0, 0.95, 0.8, 0.75$ (from left to right). 
Upper row: early stages at $t=200, 620, 400, 400$ 
(from left to right). Lower row: later stages at $t=10^4$ each.}
\label{fig2}
\end{figure}

\begin{figure}[tb]
\centering
\includegraphics[width=.8\textwidth]{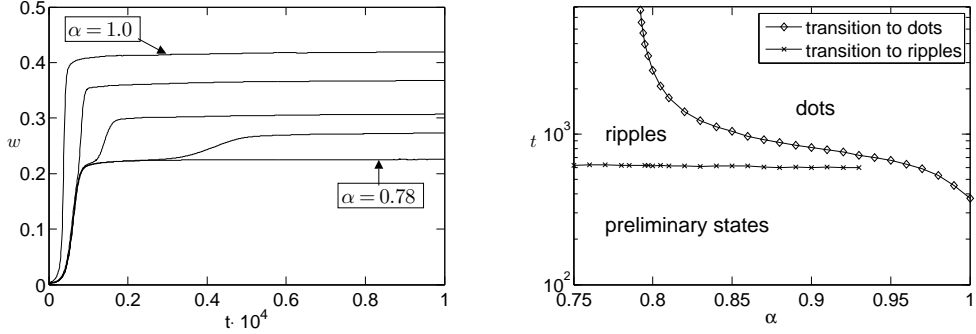} 
\caption{Left panel: Temporal evolution of the surface roughness $w$. 
for $\gamma=0.24$, $\beta=2.0$, $\alpha=1.0, 0.9, 0.82, 0.795, 0.78$ 
corresponding to the curves from top to bottom using an ensemble  
average over 40 runs. Right panel: Transition time $t_c$ to ripples 
and dots as function of $\alpha$ with fixed parameters $\gamma=0.24$ and 
$\beta=2.0$. The regions where dots or ripples exist are delimitated 
by the turning points of the surface roughness $w(t)$ as indicated 
in the left panel. For $\alpha>0.93$ only one turning point can be 
distinguished.}
\label{fig3}
\end{figure}

\begin{figure}
\centering
\includegraphics[width=.9\textwidth]{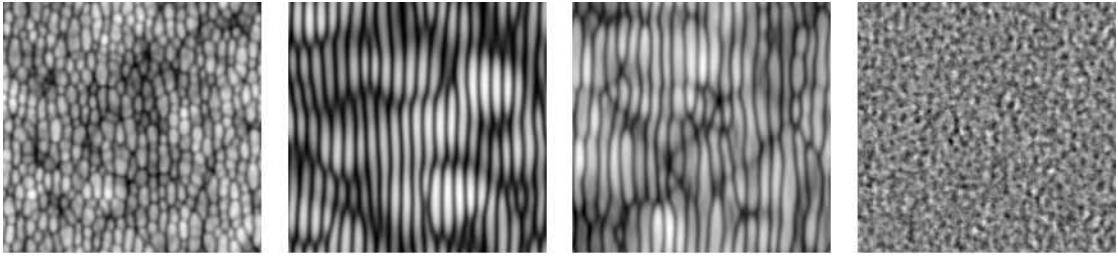}
\caption{Influence of the damping $\gamma$ on the surface morphology. 
From left to right: low damping $\gamma=0.06$ for a late stage ($t=10^4$) at $\alpha=0.8$, $\beta=2.0$, 
 an early stage ($t=50$) at $\alpha=-0.8$, $\beta=2.0$, and corresponding late stage
($t=10^4$); strong damping $\gamma=0.3$ for a late stage ($t=10^4$) at $\alpha=0.8$, $\beta=2.0$.
Surface roughness for these four parameter constellations averaged over 
$40$ runs, left to right:
 $w  = 1.25, 1.02, 1.16, 1.79\cdot 10^{-3}$.}
\label{fig4}
\end{figure}

{\it Transition from ripples to dots. -} To numerically investigate how 
hexagonally arranged dot patterns can 
turn into ripples, we fix the anisotropy of the nonlinear term in (\ref{adks1}) 
to a representative value $\beta=2$ and the damping $\gamma$ to a value in the  
unstable regime of $h_{FF}=0$ but close to the instability, $\gamma=0.24$, and focus on
$\alpha \leq 1$. 
Numerical simulations have been performed using a finite difference 
method with periodic boundary conditions, a mesh size $200\times 200$, spatial
step size $dx=1$, temporal step size $dt=0.01$,  an initially flat surface 
and an amplitude of the Gaussian white noise $A=5\times 10^{-4}$ corresponding to 
a noise covariance $D=4.1\overline{6} \times10^{-6}$. 
Varying  the anisotropy $\alpha$  off from its isotropic limit $\alpha=1$ in  
simulations of (\ref{adks1}), the following 
scenario as shown in Fig. \ref{fig2} can be observed: 
(i) for $\alpha=1$ the initially stochastically rough
surface first evolves into a worm-like structures that turn at later stages 
in a hexagonally arranged pattern with some defects.   
(ii) Lowering $\alpha$ to a value of $0.95$, there is a drastic change in the
preliminary stage of the evolution; slightly deformed n-ripples with some
superimposed deformation structure in the $y$-direction occur. At later stages,
however, the pattern stabilizes into a rather regular hexagonally arranged 
structure with 
even fewer defects than in (i). Consequently, even {\it for $\alpha\neq 1$ and 
$\beta\neq 1$ hexagonally arranged dot patterns can be observed}. 
Note, however, that these 
structures are not perfectly hexagonal since, as a result of any anistropy, the 
amplitudes of the hexagonal dots are no longer identical.  
(iii) Lowering $\alpha$ further to a value of $0.8$, the preliminary stages are
n-ripples with only minor modulations; surprisingly, the later stages consist 
of a highly regular arrangement of dots, by far more pronounced than in 
the case (i) or (ii). In our simulations, the evolution to the dot structure 
occurs via a successive periodic lacing up of the original n-ripples in 
the $y$-direction until a basically hexagonally arranged structure being 
slightly elongated
in the $y$-direction is reached. 
(iv) Lowering $\alpha$ to a value of $0.75$, the preliminary stages consisting of 
slightly modulated n-ripples stabilize for later stages to a basically perfect
n-ripple state implying that a transition from 
long-time dot patterns to long-time ripple patterns occurs 
in the range $0.78<\alpha<0.8$. \\
To investigate this transition quantitatively, we found that the  
temporal evolution of the surface roughness,  
$w(t)= \langle\overline{(h-\overline{h})^2}\rangle^{1/2}$ with 
the overbar  and $\langle...\rangle$ denoting the spatial and 
ensemble average,  
shows that the cases (ii) and (iii) are quite distinct from the other 
two cases.  This is because $w(t)$ exhibits a double plateau behavior 
reflecting the transient ripple and the long-time dot structures 
with a turning point between these plateaus, as shown in the left 
panel of Fig. \ref{fig3}. Therefore, this turning point can be considered 
as an indicator for the evolution to long-time dot structures; 
its temporal position increases
with lowering $\alpha$ until it vanishes at infinity. 
To substantiate this, we study the dependence of this turning point $t_c$, 
depicted by lozenges in the right part of Fig. \ref{fig3}, 
as function of $\alpha$.  Also marked are the regions where 
preliminary states distinct from almost perfect ripples occur.  
As can be read off from this figure, these preliminary 
states generically develop into n-ripple structures followed subsequently by 
a transition to long-time dot structures if $\alpha<0.93$. For values of 
$\alpha$ smaller than about $0.78$ only  n-ripple structures are observable. 
This signals a bifurcation between long-time ripples 
and long-time dot structures at that value of $\alpha$. 
To elucidate these results, this transition is not just restricted to
the close vicinity of normal to oblique angle of incidence;   
it can also occur when the ratio $\alpha=a_{1y}/a_{1x}$ is accidentally 
close to unity for more oblique angles of incidence and, consequently, 
also hexagonally ordered dot structures show up in such parameter ranges. 
So far, the transition from ripples to dots, $\alpha_T$ has been obtained for  
a fixed value of $\gamma$. Numerical tests indicate that the transition line 
$\alpha_T(\gamma)$ merges into the linear instability $\gamma=0.25$ as 
$\alpha \rightarrow 1$. 
Finally, as a result of the afore-mentioned property (A) of (\ref{adks1}), 
an analogous scenario happens close the instability of $h_{FF}=0$ for
$\alpha>0$ with the roles of n- and t-ripples being exchanged.
\\ 
{\it Effect of damping. -} To understand the role of the non-local term 
in (\ref{adksnl}) or the damping in (\ref{adks1}), respectively, we show 
as a representative example in the first panel from the left
of Fig.\ref{fig4} 
late stages of the surface morphology for the same values as in the 
third row of Fig. \ref{fig2}, except that the damping parameter has been
drastically reduced to a value of $\gamma=0.06$. Obviously, lowering $\gamma$ 
has the effect that the rather periodic hexagonal dot structure seen 
for $\gamma=0.24$ has been coarsened into a pattern 
consisting of irregularly sized, droplet-like structures with a
preferential elongation in $y$-direction. Flipping the sign of $\alpha$ 
for this parameter
constellation, as shown in the second and third panel of Fig.
\ref{fig4}, the initial and late stages of the surface evolution exhibit 
strong, irregular superimposed modulations of the n-ripples in the 
$y$-direction.  Consequently, the damping $\gamma$ acts as a order-disorder 
parameter, i.e. it gradually changes the pattern from the maximally disordered limit 
for $\gamma=0$, i.e. the aKSE limit,  to a quite regular ripple or dot pattern 
close to $\gamma=0.25$ (if $\alpha<1$). For even larger $\gamma$, the  
deterministic limit of the adKSE possesses a linearly stable flat surface. 
Due to the nonzero stochasticity $\eta$  in our simulations, however, 
the resulting pattern for such values of $\gamma$ is stochastically rough, as 
depicted in the forth panel of Fig.\ref{fig4}. 
Most remarkably,   
the corresponding surface roughness is drastically reduced (by
almost three orders of magnitude) in comparison to the cases with $\gamma\leq 0.24$.  
Therefore, the pattern for $\gamma>1/4$, can be interpreted as a basically smooth  
surface with some stochastic variations that reflect the stochastic component 
of the sputter process. Such patterns have recently been reported by 
Frost et al. \cite{frost-smooth}. As a consequence, transitions to such 
basically smooth  patterns triggered by changing the angle 
of incidence are preluded by a quite regular ripple or dot pattern. 
Assuming a nonlinear dependence of $\alpha$ and $\gamma$ on the angle of
incidence and keeping $\beta$ fixed (cf. also Fig.\ref{fig1}), a variety of 
distinct transitions between patterns such as  n-ripples $\leftrightarrow$ 
smooth surfaces $\leftrightarrow$ t- or n-ripples, n-ripples 
$\leftrightarrow$ dot structures $\leftrightarrow$ t- or n-ripples 
as well as more disordered states seem to be possible. Some of these transitions 
have been experimentally reported in \cite{ziberi,ziberi2,allmers}
 (cf. also our introductory part). 
Using property (B) of Eq. (\ref{adksnn}), we also infer that 
patterns that look like  negative images  (e.g. mounds exchanged by holes)  of 
the ones in Fig.\ref{fig4} can be seen by going to 
negative values of $a_{3x}$ and $a_{3y}$.
Finally, our numerical simulations show that the transition to 
hexagonally arranged dots is generically subcritical  
implying that starting from an appropriately 
prestructured surface, dot patterns might evolve even for 
$\gamma>1/4$. 

{\it Cancellation modes (CM). -} Straight ripples oriented obliquely 
to the $x$-direction, as found by Rost and Krug \cite{rost} in the aKSE, 
can also be recovered  
in the adKSE. Assuming solutions $h(x,y,t)=f(x- uy,t)$ that are constant 
along lines $x=s+uy$ and demanding that the nonlinearity in the adKSE  
vanishes, implies  $u=\pm\sqrt{-1/\beta}$. Their evolution governed by 
$\partial_t f =  -(\gamma+ 1+  \alpha u^2) \partial^2_s f - 
(1+u^2)^2 \partial^4_s f$ has the maximum growth rate 
 $\sigma_{max}=-\gamma+(\beta-\alpha)^2/[4(1-\beta)^2]$ 
at a wave number $q=\sqrt{\beta(\beta-\alpha)/2(\beta-1)^2}$. 
Consequently, CM exist only if $\beta<0$ and $\alpha>\beta$ as in the case  
of the aKSE \cite{rost}. Remarkably, CM exist as stationary solutions 
if $\gamma= (\beta-\alpha)^2/[4(1-\beta)^2]$ in contrast to the 
aKSE \cite{rost}. \\
{\it Conclusions. -} The adKSE (\ref{adks1}) presented here 
seems to be a promissing candidate 
for a continuum model for low energy ion-beam erosion under oblique ion incidence 
since it can reproduce many distinct types of patterns also seen in experiments. 
To substantiate this, comparison with so far not available detailed experimental 
studies is needed.
A more detailed discussion of the deterministic adKSE focussing on the 
bifurcation structure of its patterns will be given elsewhere \cite{reh1}.

\end{document}